\documentclass[conference]{IEEEtran}
\IEEEoverridecommandlockouts
\usepackage{cite}
\usepackage{amsmath,amssymb,amsfonts}
\usepackage{algorithmic}
\usepackage{graphicx}
\usepackage{textcomp}
\usepackage{xcolor}
\usepackage{booktabs} 
\usepackage{siunitx}  
\usepackage{caption}  
\usepackage{bm}

\def\BibTeX{{\rm B\kern-.05em{\sc i\kern-.025em b}\kern-.08em
    T\kern-.1667em\lower.7ex\hbox{E}\kern-.125emX}}
\begin{document}

\title{Neuro-Informed Joint Learning Enhances Cognitive Workload Decoding in Portable BCIs}

\author{
\IEEEauthorblockN{Xiaoxiao Yang}
\IEEEauthorblockA{\textit{Peiyang Brain-Computer Interface \&} \\ 
\textit{Smart Health Inst.} \\ 
Xiamen, China \\ 
xiaoxiao.yang@ieee.org}
\and
\IEEEauthorblockN{Chao Feng}
\IEEEauthorblockA{\textit{Peiyang Brain-Computer Interface \&} \\
\textit{Smart Health Inst.} \\ 
Xiamen, China \\ 
charles\_feng@intretech.com}
\and
\IEEEauthorblockN{Jiancheng Chen}
\IEEEauthorblockA{\textit{Peiyang Brain-Computer Interface \&} \\
\textit{Smart Health Inst.} \\
Xiamen, China \\ 
xmcjc@intretech.com}
}
\maketitle

\begin{abstract}
Portable and wearable consumer-grade electroencephalography (EEG) devices, like Muse headbands, offer unprecedented mobility for daily brain-computer interface (BCI) applications, including cognitive load detection. However, the exacerbated non-stationarity in portable EEG signals constrains data fidelity and decoding accuracy, creating a fundamental trade-off between portability and performance. To mitigate such limitation, we propose MuseCogNet (Muse-based Cognitive Network), a unified joint learning framework integrating self-supervised and supervised training paradigms. In particular, we introduce an EEG-grounded self-supervised reconstruction loss based on average pooling to capture robust neurophysiological patterns, while cross-entropy loss refines task-specific cognitive discriminants. This joint learning framework resembles the bottom-up and top-down attention in humans, enabling MuseCogNet to significantly outperform state-of-the-art methods on a publicly available Muse dataset and establish an implementable pathway for neurocognitive monitoring in ecological settings.
\end{abstract}

\begin{IEEEkeywords}
BCI, EEG, Self-Supervised Learning, Cognitive Load, Wearable, Human Attention
\end{IEEEkeywords}

\section{Introduction}
Overloaded cognitive states in critical scenarios like driving are major contributors to car accidents~\cite{neng-chao_research_2018}, necessitating real-time cognitive workload detection. EEG-based BCIs offer promise due to non-invasiveness and temporal resolution. Yet researches based on medical-grade EEG devices~\cite{yiu_keeping_2025, lin_hatnet_2025} face deployment barriers: their complex setups require expertise and continuous monitoring. Consumer-grade devices like Muse headbands provide accessibility but introduce amplified non-stationarity and signal variability from mobile operation. This demands robust algorithms to identify neural signatures amid environmental noise to enhance decoding accuracy. 
\textbf{Self-supervised learning} (SSL) provides a principled framework for learning noise-invariant representations directly from unlabeled EEG. By designing pretext tasks that exploit inherent electrophysiological pattern~\cite{weng_self-supervised_2025}, SSL models can discover neural signatures resilient to environmental interference. However, traditional voltage reconstruction objective has been reported to underperform due to inherent signal instability~\cite{jiang2024large}. We propose that reconstructing temporally aggregated (average-pooled) representations is more physiologically meaningful. Specifically, by forcing the model to reconstruct the average-pooled temporal segments, we argue that the model can integrate neuroinformative insights while attenuating transient noise, thus capturing sustained neural processes underlying cognitive states. For downstream classification, we integrate \textbf{supervised learning} (SL) to complement SSL's physiological grounding. While SSL reveals fundamental neural dynamics, SL optimizes discriminative boundaries by amplifying task-specific domain-relevant representations. The SSL-SL synergy forms a \textbf{computational duality}: SSL's bottom-up feature extraction and SL's top-down decision refinement jointly emulate the human neurocognitive processing stream, where sensory-level representations are progressively modulated by task-specific demands.

The main contribution of this paper is summarized below:

\textbullet We propose a novel SSL reconstruction objective based on average-pooled EEG, adapting to EEG's intrinsic non-stationary nature.

\textbullet We implement the first joint SSL-SL framework deployed on wearable mobile EEG headsets (Muse), establishing a portable paradigm for ecological cognitive load monitoring.

\section{Methodology} 

\textbf{Encoder}: We deploy parallel convolutions with increasing kernel lengths $(1\times16)$, $(1\times32)$, $(1\times64)$, $(1\times128)$ to capture spectral features $>$16Hz, 8Hz, 4Hz, and 2Hz (since sampling rate is 256Hz). Concatenated features undergo batch normalization and ELU activation, followed by dual temporal pooling (variance + average)~\cite{ma2024attention}. Pooled streams merge via $(2\times1)$ convolution to form latent representations.

\textbf{Dual-Path Decoding}: 
- \textit{SSL Path}: Reconstructs average-pooled signals via two linear layers
- \textit{SL Path}: Projects latent features to logits via 1D convolution

\textbf{Losses}: 
Losses are tailored individually for SSL and SL components. The \textbf{self-supervised} reconstruction loss $\mathcal{L}_{\text{pool}}$ measures the difference between true average pooled EEG $\bm{X}_{\text{avg\_pooled}}$ and reconstructed average pooled EEG $\hat{\bm{X}}_{\text{avg\_pooled}}$ to force model to learn temporally robust features, easing the adversial effect of non-stationarity in EEG. For a given EEG segment, the average-pooled loss is defined as below:
\begin{equation}
\mathcal{L}_{\text{pool}} = \frac{1}{C \times T'} \sum_{c=1}^{C} \sum_{t=1}^{T'} \left( X_{\text{avg\_pooled}}^{(c,t)} - \hat{X}_{\text{avg\_pooled}}^{(c,t)} \right)^2
\end{equation}
where $C$ is the number of channels, $T'$ is the pooled dimension of EEG signals.

The \textbf{supervised} cross-entropy loss refines the features into task-specific representations. For any given EEG segment, this is defined by:
\begin{equation}
\mathcal{L}_{\text{CE}} = -\sum_{k=1}^{K} y_{k} \log\left( p_{k} \right)
\end{equation}
where $K$ indicates the number of possible categories (three in this case, namely low, medium or high level of cognitive load), and $p$ is computed via softmax normalization of the SL decoded logits.

The \textbf{overall loss} for a given sample is calculated as follows:
\begin{equation}
\mathcal{L} =  \mathcal{L}_{\text{CE}} + \lambda \mathcal{L}_{\text{pool}} 
\end{equation}
where $\lambda$ is the SSL loss strength, which is set to 0.5 empirically.

\section{Experiment and Analysis} 

\subsection{Dataset and Experimental Setup}
We utilize the CL-Drive dataset~\cite{muse_cldrive} captured using Muse S headbands during driving simulations. Reported cognitive load scores from 21 participants are categorized into 3 levels (low/medium/high) for each 10-second intervals. To maximize label-temporal relevance, models received the final 2-second segment of each interval. 

EEG data is bandpass-filtered (0.1-75 Hz) with 60Hz notch filtering. Training employed AdamW (lr=0.001) with 100 epochs and batch size 64. Evaluation follows leave-one-subject-out (LOSO) protocol: each participant served as test set once, mimicking real-world new-user scenarios. Reported metrics are averaged performance across all test subjects.

\begin{table}[htbp]
  \centering
  \caption{Comparison of LOSO Performance across Models}
  \label{tab:model_comparison}
  \begin{tabular}{
    l              
    S[table-format=2.2] 
    S[table-format=2.2]
    S[table-format=2.2]
    S[table-format=2.2]
  }
    \toprule
    \textbf{Models}    & \textbf{EEGNet} & \textbf{Deep ConvNet} & \textbf{ResNet} & \textbf{MuseCogNet} \\
    \midrule
    Accuracy (\%) & 57.07 & 57.07 & 58.41 & \textbf{62.68} \\
    F1 score (\%) & 52.09 & 51.25 & 51.61 & \textbf{57.24} \\
    \bottomrule
  \end{tabular}
  \vspace{0.5em} 

\end{table}

\begin{table}[htbp]
  \centering
  \caption{Ablation Results}
  \label{tab:model_ablation}
  \begin{tabular}{
    l              
    S[table-format=2.2] 
    S[table-format=2.2]
    S[table-format=2.2]
    S[table-format=2.2]
  }
    \toprule
    \textbf{Models}     & \textbf{W/O SSL} & \textbf{MuseCogNet}\\
    \midrule
    Accuracy (\%) & 60.77 & \textbf{62.68}\\
    F1 score (\%) & 54.60 & \textbf{57.24}\\
    \bottomrule
  \end{tabular}
  \vspace{0.5em} 

\end{table}

\subsection{Analysis and Discussion}
The comparative analysis reveals distinct performance advantages of our proposed MuseCogNet. As seen in Table~\ref{tab:model_comparison}, established models like EEGNet~\cite{lawhernEEGNetCompactConvolutional2018} and Deep ConvNet~\cite{schirrmeisterDeepLearningConvolutional2018} demonstrate competent baseline performance (Accuracy: 57.07\%, F1: 51.25-52.09\%), validating their status as canonical architectures for EEG decoding. Meanwhile, our implementation of ResNet achieves 58.41\% accuracy, closely matching the original CL-Drive benchmark(58.13\%)~\cite{muse_cldrive}, confirming proper experimental replication. Critically, MuseCogNet achieves superior performance of 62.68\% accuracy and 57.24\% F1, representing \textbf{absolute gains of 4.27\% accuracy and 5.63\% F1} over the strongest baseline (ResNet). We attribute this significant margin to our novel joint learning paradigm enhanced by average-pooled reconstruction objective, where the SSL component learns steady neurodynamic patterns that conventional supervised CNNs fail to capture. The ablation study (Table~\ref{tab:model_ablation}) provides definitive evidence: Removing SSL components causes \textbf{performance degradation of 1.91\% accuracy and 2.64\% F1}, confirming that our average-pooled reconstruction objective contributes substantially to feature robustness. 

Overall, MuseCogNet enables reliable cognitive load monitoring on portable EEG by leveraging SSL-SL synergy—where SSL's bottom-up feature extraction and SL's top-down discriminant refinement exhibit functional similarities to cortical sensory-cognitive processing hierarchies. This joint processing architecture not only achieves superior performance but also suggests intriguing parallels to human attentional mechanisms worthy of future exploration.

\bibliographystyle{IEEEtran}
\bibliography{IEEEabrv,citations}

\end{document}